\documentclass[
twocolumn,
aps,
superscriptaddress, 
amsmath,
amssymb,
amsfonts,
floatfix,
square
]{revtex4-1}
\usepackage[colorlinks=true, allcolors=blue]{hyperref}
\usepackage{bm}
\usepackage{lipsum}
\usepackage[utf8]{inputenc}
\usepackage[nolist]{acronym}
\usepackage{todonotes}
\usepackage[caption=false]{subfig}
\usepackage{graphicx}
\usepackage{siunitx}
\usepackage{import}
\usepackage{float}
\setcitestyle{numbers,square}

\newcommand*\diff{\mathop{}\!\mathrm{d}}

\newcommand{\vect}[1]{\boldsymbol{\mathbf{#1}}}
\renewcommand{\Re}{\operatorname{Re}}

\DeclareMathOperator{\reg}{reg}

\renewcommand{\v}[1]{{\ensuremath{\boldsymbol{\mathbf{#1}}}}}

\begin{document}
\title{First order superconducting phase transition in chiral $p+ip$ system}
\author{H\aa vard Homleid Haugen}
\affiliation{\footnotesize Department of Physics, Norwegian University of Science and Technology, NO-7491, Trondheim, Norway}
\affiliation{\footnotesize Center for Quantum Spintronics, Department of Physics, Norwegian University of Science and Technology, NO-7491, Trondheim, Norway}
\author{Egor Babaev}
\affiliation{Department of Physics, KTH-Royal Institute of Technology, Stockholm SE-10691, Sweden}
\author{Fredrik Nicolai Krohg}
\author{Asle Sudb\o}
\affiliation{\footnotesize Department of Physics, Norwegian University of Science and Technology, NO-7491, Trondheim, Norway}
\affiliation{\footnotesize Center for Quantum Spintronics, Department of Physics, Norwegian University of Science and Technology, NO-7491, Trondheim, Norway}
\date{\today}

\begin{abstract}
We use large-scale Monte Carlo computations to study the phase transitions between a two-component chiral $p$-wave superconductor and a normal state in zero   external magnetic field.
We find  a first order phase transition
from the normal state to a chiral superconducting state, due to interplay between vortices and domain walls.  
\end{abstract}
\maketitle

\section{Introduction} \label{sec:Introduction}

\begin{acronym}
  \acro{TRS}{time reversal symmetry}
  \acro{GL}{Ginzburg Landau}
  \acro{MGT}{mixed gradient terms}
  \acro{MCS}{Monte-Carlo sweep}
\end{acronym}

Chiral superconductors constitute a class of unconventional superconductors whose order parameter features finite angular momentum and a phase that winds around the Fermi surface \cite{Kallin_2016}. The chiral nature arises from spontaneously broken \ac{TRS}, which yields a two-fold degenerate superconducting state with broken \(U(1)\times Z_2\)-symmetry. Chiral superconductors are of fundamental interest because they are predicted to display topological properties such as Majorana modes in vortex cores and edge currents leading to a quantized thermal Hall conductance \cite{read2000paired, PhysRevB.69.184511, PhysRevB.60.4245, PhysRevB.93.024510}. 

The prototypical chiral \(p\)-wave superfluid state is realized in \(A\)-phase of superfluid \(\phantom{}^{3}\mathrm{He}\) \cite{PhysRevLett.5.136, PhysRevLett.30.1108, PhysRev.131.1553}.
The search for chiral \(p\)-wave pairing in a bulk superconductor has been going on since the discovery of superfluid \(\phantom{}^{3}\mathrm{He}\). For many years, the leading candidate has been the extensively investigated superconductor \(\mathrm{Sr}_2\mathrm{RuO}_4\); a highly anisotropic layered material with tetragonal crystal structure and strong spin-orbit coupling \cite{maeno1994superconductivity, luke1998time, xia2006high, jang2011observation, mackenzie2017even}. ARPES-measurements have revealed three bands crossing the Fermi-surface, supporting a multi-component theory \cite{PhysRevLett.85.5194}. Several groups have also found that in zero field there is a single phase transition, where \ac{TRS} is broken along with the onset of superconductivity \cite{xia2006high, luke1998time, nishizaki1997pairing,grinenko2021unsplit}, while split transitions were  reported to arise under  strain \cite{grinenko2020split}. 
However the evidence against the chiral \(p\)-wave superconductivity has been growing in recent years. The first notable example was  the absence of chiral edge currents that should produce magnetic signatures at the boundary between domains of opposite chirality \cite{PhysRevB.89.144504, kirtley2007upper,hicks2010limits}. Recently, the mounting evidence against the chiral p-wave pairing lead to the discussion of other order parameters in an attempt to reconcile all the experimental data, such as near-degenerate between \(d\)- and \(g\)-wave pairing for \(\mathrm{Sr}_2\mathrm{RuO}_4\) \cite{romer2019knight, kivelson2020proposal,PhysRevResearch.2.032023,PhysRevB.100.104501,PhysRevB.94.104501}. Recent studies of ultrasound
\cite{ghosh2021thermodynamic,benhabib2021ultrasound}, and vortex state \cite{ray2014muon} point to a multi-component order parameter.

Another candidate for chiral triplet superconductivity is the  heavy fermion superconductor \(\mathrm{UPt}_3\) \cite{joynt2002superconducting, strand2009evidence, schemm2014observation}. Unlike \(\mathrm{Sr}_2\mathrm{RuO}_4\), it is claimed to feature two separate phase transitions in zero applied magnetic field, where \ac{TRS} is spontaneously broken within the superconducting phase \cite{PhysRevLett.62.1411, adenwalla1990phase, avers2020broken}. The superconducting state in \(\mathrm{UPt}_3\) is believed to be chiral \(f\)-wave with an order parameter that has the two-dimensional irreducible representation \(E_{2u}\) \cite{sauls1994order}. Although this is a higher order pairing than chiral \(p\)-wave, our theoretical description will be relevant for \(\mathrm{UPt}_3\) since the order parameter symmetry group has the same irreducible representation. In more recent works, chiral superconductivity has also been claimed in other systems, such as Van der Waals materials and nano tubes \cite{ribak2020chiral, qin2017superconductivity, jiao2020chiral}.

Even after decades of research, the nature of multi-component superconductivity in \(\mathrm{Sr}_2\mathrm{RuO}_4\) remains a puzzle.
This fact and the emergence of new candidates for chiral superconductors raise the need to understand the nature of superconducting phase transition in a chiral $p$-wave superconductor beyond mean-field approximations and possible clues it may yield in real materials.

The question of fluctuations in a chiral \(p\)-wave superconductor is nontrivial because it breaks two symmetries: $U(1)$ and $Z_2$.
Therefore, in general, fluctuations can cause a single transition or a sequence of transitions. 
A similar question arises for $s+is$ superconductors, that shares the $U(1)\times Z_2$ symmetry and has been studied by numerical methods \cite{Bojesen2013,Bojesen2014}. Recent experiments reported fluctuations-induced splitting of the phase transition \cite{grinenko2021bosonic}.
Analogous questions for chiral p-wave superconductors were studied in \cite{fischer2016fluctuation}, but no Monte-Carlo calculations were performed for this problem.
In this paper, we use large-scale Monte-Carlo calculations to study the phase transition  a chiral two-component superconductor transition in \ac{GL}-theory for an \(E_{2u}\) order parameter.
Before we proceed to calculations, we note that the problem is related to the more general question of the  phase transitions in multi-component gauge theories, where large-scale Monte-Carlo studies were performed.
For a \(U(1) \times U(1)\) two-component London superconductor, it has been shown that for moderate values of the gauge charge and equal amplitudes in the two ordering fields there is a single first order phase transition where both symmetries are broken at the same temperature. For high values of the gauge charge the single transition line splits into two separate transitions predicting an intermediate metallic superfluid with broken global \(U(1)\) symmetry but restored local \(U(1)\) symmetry \cite{PhysRevB.82.134511,PhysRevB.71.214509,kuklov2006deconfined,babaev2004phase,babaev2004superconductor}. In Refs.~\cite{dahl2008preemptive, PhysRevB.82.134511} the merging of the two phase transitions was coined a preemptive phase-transition, where ordering in one symmetry sector of the model leads to ordering in the other. For the case of interacting  $U(1)\times U(1) $ neutral superfluid a detailed study of the first order character of the phase transition was presented in \cite{kuklov2006deconfined}, where also the existence of a tricritical point was reported. 
Similarly, for a \(SU(2)\)-symmetric model, where the amplitudes of the two matter fields $(\Psi_1,\Psi_2)$ are related by a CP\(^1\) constraint $|\Psi_1|^2+|\Psi_2|^2=1$, a single transition was found for moderate values of the gauge charge, which split into two transitions for higher values \cite{kuklov2008deconfined,PhysRevB.87.134503}. The model we consider in this paper 
is different  from a  \(U(1)\times U(1)\) London superconductor, due to the presence of a term that explicitly breaks the global \(U(1)\) symmetry down to a \(Z_2\)-symmetry.
It is also different from $s+is$ superconductor due to the structure of a so-called \ac{MGT}. These terms are products of two gradient terms, as in the standard kinetic energy, but where the two factors are gradients in different directions involving different order-parameter components (see below). Such terms are common for chiral p-wave superconductors \cite{agterberg1998vortex} and can also originate for instance with spin-orbit coupling \cite{PhysRevB.98.014510}. Such terms will provide  an additional direct coupling between the $U(1)$- and $Z_2$-symmetry sectors of the model.

\section{Model}\label{sec:Model}
\subsection{Ginzburg Landau model}\label{subsec:gl_model}
We consider a superconductor with tetragonal crystal structure and spin orbit coupling, belonging to the point group \(D_{4h}\). Gauge invariance and \ac{TRS} yields the full symmetry group of the system \(\mathcal{G} = D_{4h} \times U(1) \times Z_2\). In the two-dimensional odd-parity representation \(E_{2u}\), the superconducting gap function may be written as \(\vect{d}(\vect{k}) = (\eta_x k_x + \eta_y k_y)\hat{z}\). The complex matter fields (components) describe two types of Cooper pairs in the theory, and can be written in terms of an amplitude and a phase on the form \(\eta_i = \rho_i e^{i\theta_i}\). This leads to a \ac{GL} energy functional \(E = \int f \diff^3 r\) where the dimensionless energy density is given by \cite{agterberg1998vortex, RevModPhys.63.239, PhysRevB.94.104509}
\begin{subequations}
\begin{alignat}{1}
f =& -\alpha(|\eta_x|^2 + |\eta_y|^2) + \frac{u_0}{2}\left(|\eta_x|^4 + |\eta_y|^4\right) \label{eq:GL_pot} \\
    & + \gamma|\eta_x\eta_y|^2\cos 2(\theta_x-\theta_y) \nonumber \\
    & + |\vect{D}\eta_x|^2 + |\vect{D}\eta_y|^2 + |\nabla \times \vect{A}|^2  \label{eq:GL_kin+B}\\
    & + \gamma_m\left[(D_x\eta_x)(D_y\eta_y)^{*}+(D_y\eta_x)(D_x\eta_y)^{*} + \mathrm{h.c.}\right]. \label{eq:GL_mgt}
\end{alignat}\label{eq:GL_full}
\end{subequations}
The matter fields are minimally coupled to the gauge field \(\vect{A}\) through covariant derivatives \(\vect{D} = \nabla - ig\vect{A}\) and the energy is normalized to the condensation energy \(f_0 = B_c^2/4\pi\), where \(B_c\) is the critical magnetic field. Lengths are given in units of \(\xi = 1/\sqrt{\alpha}\). \(\alpha\) and \(u_0\) set the scale for the critical temperature. Our results do not sensitively depend on their precise values, but they apply only in a domain where the phase-only approximation is approximately valid, for example deep in the type-II regime. The coefficients \(\gamma\) and \(\gamma_m\), which control the strength of the intra-component potential and \ac{MGT} respectively, will typically be sensitive the electronic structure and in particular the shape of the Fermi surface \cite{agterberg1998vortex,PhysRevB.98.014510}. In order to present a systematic study of the effects these terms have on a class of physical systems, we will treat them as phenomenological parameters. They are still constrained by our numerical methods, discussed further in Sec.~\ref{subsec:London_limit}. The decay of  magnetic fields in this model usually involves multiple modes and multiple length scales \cite{speight2019chiral}, that yields further differences compared to $s+is$
models previously studied in Monte-Carlo simulations \cite{Bojesen2013,Bojesen2014}. In what follows we will not distinguish between the subdominant electromagnetic scales. 

The mean field ground state of Eq.~\eqref{eq:GL_full} is found by setting \(\vect{A} = 0\) and ignoring spatial variations in the matter fields. Minimization of the potential energy in Eq.~\eqref{eq:GL_pot} then yields the ground state
\begin{equation}\label{eq:MFGS_amps}
    |\eta_x| = |\eta_y| = \sqrt{\frac{\alpha}{u_0 - \gamma}} \equiv \rho_0,
\end{equation}
\begin{equation}\label{eq:MFGS_phase}
    \theta_x - \theta_y = \pm \pi/2 \equiv \theta_0.
\end{equation}
We find two degenerate solutions due to the phase-locking term. Theses are related by a \(Z_2\) symmetry operation which will be discussed in more detail in Sec.~\ref{subsec:Charged_chiral_sym_sec}. Finally, we note that this ground state gives an order parameter on the form \(k_x \pm i k_y\), corresponding to a superconducting state with chiral \(p\)-wave pairing which spontaneously breaks the \(U(1) \times Z_2\) symmetry of the theory.

\subsection{The London limit}\label{subsec:London_limit}
In order to perform Monte Carlo simulations on the free energy introduced in Eq.~\eqref{eq:GL_full}, we will work within the London approximation where the amplitudes of the matter fields are frozen. The London limit is commonly used for similar models \cite{PhysRevB.82.134511, PhysRevB.71.214509}. However, in the case of a multi-component order parameter, and with the addition of Ising anisotropy and \ac{MGT}, such an approach requires considerable care and is generally not applicable \cite{speight2019chiral}.
 
We will first explicitly assess the validity of this approach, following a similar but not identical method to the one presented in \cite{speight2019chiral}. To this end, we expand all fluctuating fields to second order in deviations from their mean-field values, introducing 
\begin{equation}\label{eq:epsilon_i}
    \epsilon_i = \rho_i - \rho_0,
\end{equation}
\begin{equation}\label{eq:theta_delta}
    \theta_\Delta = \frac{1}{2}\left( \theta_x - \theta_y -\theta_0 \right),
\end{equation}
\begin{equation}\label{eq:p_i}
    p_i = A_i - \frac{1}{g}\partial_i\theta_\Sigma,
\end{equation}
\begin{equation}\label{eq:theta_sigma}
    \theta_\Sigma = (\theta_x + \theta_y)/2,
\end{equation}
where \(p_i\) essentially is a gauge-invariant current. Expanding the energy to second order in these fluctuations and Fourier transforming, we obtain an expression on the form
\begin{equation}\label{eq:GL_energy_matrix}
    f = f_0 + \vect{v}\vect{G}\vect{v}^{\dagger},
\end{equation}
where \(f_0\) is the ground state energy, \(\vect{G}(\vect{k})\) is a matrix describing the coupling between fluctuations in different fields and \(\vect{v}\) is the fluctuation vector given by 
\begin{equation}\label{eq:fluctuation_vector}
    \vect{v}(\vect{k}) = \begin{pmatrix}
    \epsilon_+ & \epsilon_- & \theta_\Delta & p_x & p_y
    \end{pmatrix}.
\end{equation}
We have introduced a rotated amplitude basis \(\epsilon_\pm = (\epsilon_x \pm \epsilon_y)/\sqrt{2}\) in order to simplify the structure of  the coupling matrix. The exact form of the coupling matrix, along with details of the derivation are given in Appendix \ref{app:Coupling_matrix}. 

To determine what fluctuations are most important, the coupling matrix is diagonalized to obtain the lowest eigenvalue \(\lambda_*\) along with the corresponding eigenvector \(\vect{\psi_*}\). In the absence of \ac{MGT}, the coupling matrix is already diagonal and the eigenvectors are pure modes with fluctuations in only one field.  In the long wavelength limit, we then find that the phase-difference mode \(\theta_\Delta\) corresponds to the lowest eigenvalue for low values of the Ising anisotropy 
\begin{equation}
    \gamma \leq 0.17 
\end{equation}
when \(\alpha = u_0 = g = 1.0\). Above this value amplitude fluctuations become important as the \(\epsilon_-\) mode corresponds to the lowest eigenvalue. The effect of \ac{MGT} is that the eigenvectors become mixed modes with multiple non-zero entries for non-zero momentum \cite{speight2019chiral}. To investigate the degree of mixing, we plot the \(k\)-dependence of the non-zero entries in \(\vect{\psi_*}\) in Fig.~\ref{fig:eigen_values__nu0,1mgt0,1} with the corresponding parameters without \ac{MGT} in Fig.~\ref{fig:eigen_values__nu0,1mgt0,0}. For low momentum magnitude \(k\) the phase difference mode is now weakly mixed with \(\epsilon_+\) amplitude fluctuations, but phase difference fluctuations are still dominant. 

Note that although taking a London limit eliminates some of the mixing at the level of bare model, we find below that the phase transition is first order, so in a fluctuating model the mixing should reappear at the level of a large-scale effective field theory. Otherwise, at the level of bare model, the London limit is a good approximation for the regime of small mixing. 
\begin{figure}[htb]
    \subfloat[\label{fig:eigen_values__nu0,1mgt0,0}]{
        \includegraphics[width=\linewidth]{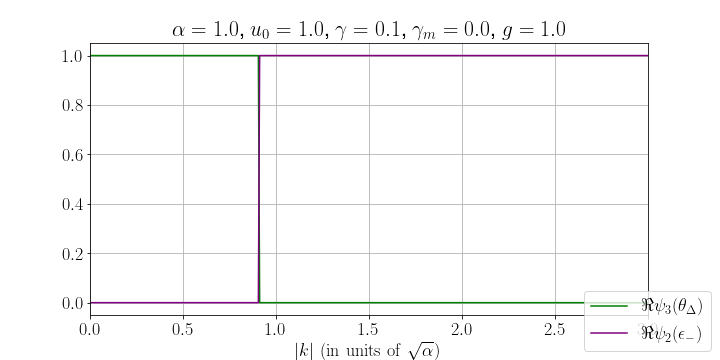}
    }
    
    \subfloat[\label{fig:eigen_values__nu0,1mgt0,1}]{
        \includegraphics[width=\linewidth]{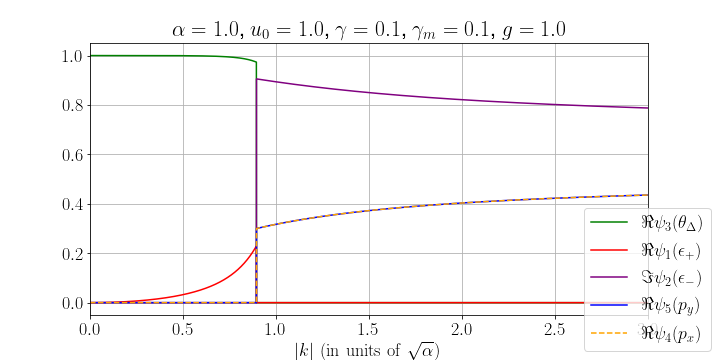}
    }
    \caption{Non-zero entries of the eigenvector \(\psi_*\) corresponding to the lowest eigenvalue of the coupling matrix, plotted along the line \(k_x = k_y\). In (a) there are no \ac{MGT} and the coupling matrix is diagonal with pure modes. In (b) we have included \ac{MGT}, which cause mixed modes with fluctuations in multiple fields. In both cases fluctuations in the phase difference are dominant in the long wavelength limit.}
    \label{fig:eigenvectors}
\end{figure}

\subsection{Charged and chiral symmetry sectors}\label{subsec:Charged_chiral_sym_sec}
In this section we introduce the chiral basis, which is obtained by a unitary transformation \(\eta_\pm = (\eta_x \pm i\eta_y)/\sqrt{2}\). Under \ac{TRS}, the chiral components transform as \(\hat{\mathcal{K}}\eta_\pm = \eta_\mp^*\). It is common to recast the model in terms of these chiral components \cite{PhysRevB.58.14484, PhysRevB.94.104509, PhysRevB.59.7076}, but in the present setting we introduce them because they provide an order parameter in the \(Z_2\) symmetry sector. If we calculate the chiral component amplitudes in terms of the \(xy\)-components, we find
\begin{equation}\label{eq:chiral_amplitudes}
    |\eta_\pm| = \sqrt{\frac{1}{2}\left[|\eta_x|^2 + |\eta_y|^2 \pm 2|\eta_x\eta_y|\sin(\theta_x - \theta_y)\right]}.
\end{equation}
By inserting the ground state values form Eqs.~\eqref{eq:MFGS_amps} and \eqref{eq:MFGS_phase}, we see that one of the chiral amplitudes is spontaneously chosen. Coming from the low-temperature regime, chiral symmetry is then restored by a proliferation of topological defects in the form of Ising domain walls separating areas of opposite chirality. From Eq.~\eqref{eq:chiral_amplitudes} we see that these domain walls can be described by a gradient in the phase-difference of the \(xy\)-components. 

The superconducting phase transition is associated with spontaneous symmetry breaking of the local \(U(1)\)-symmetry. The low temperature phase is well understood at mean-field level, where the gauge field \(\vect{A}\) acquires a mass, yielding a Meissner-effect.
  In the context of  single-component superconductors it has been shown that going beyond mean-field, the (non-local) order parameter of the \(U(1)\)-sector is still the gauge field mass, which now corresponds to the inverse magnetic penetration length of the problem. Upon heating the system, the mass of the gauge field is eventually destroyed at some critical temperature. The phase transition is driven by a proliferation of thermally excited topological defects in the form of charged vortex-loops \cite{PhysRevLett.47.1556, PhysRevB.60.15307}. 

In the London limit, we can perform a separation of variables to rewrite the model in terms of charged and chiral terms
\begin{widetext}
\begin{alignat}{1}
    f = &\frac{\rho_0^2}{2}\left[\nabla(\theta_x + \theta_y) - 2g\vect{A}\right]^2 + \frac{\rho_0^2}{2}\left[\nabla(\theta_x - \theta_y)\right]^2 + \gamma\rho_0^4\cos 2(\theta_x - \theta_y) +  |\nabla \times \vect{A}|^2 \label{eq:Sep_of_variables} \\
    & + \gamma_m \rho_0^2\cos (\theta_x-\theta_y)\left\{[\partial_x(\theta_x+\theta_y) - 2gA_x][\partial_y(\theta_x+\theta_y) - 2gA_y] - [\partial_x(\theta_x-\theta_y)][\partial_y(\theta_x-\theta_y)]\right\}. \nonumber
\end{alignat}
\end{widetext}
This form highlights the interplay between the symmetry sectors of the model in an intuitive way. We have the charged sector given by the phase-sum coupling to a gauge field with strength \(2g\). The chiral sector is governed by the phase-difference, where we have a 3D XY-model with an easy axis anisotropy that demotes the symmetry from global \(U(1)\) down to \(Z_2\). Then finally there are the \ac{MGT} that provide an explicit coupling between the two sectors. Note that even in the absence of \ac{MGT}, the two symmetry sectors are still connected as the phase-sum and phase-difference are not independent variables. 
\section{Monte Carlo simulations}\label{sec:MC_sim}
The critical properties of the model in Eq.~\eqref{eq:GL_full} in the London limit are investigated using Monte Carlo simulations. This is achieved by discretizing the model on a numerical cubic lattice, where the matter-fields live on lattice points and the gauge field is discretized through renormalized non-compact link-variables  \cite{Shimizu12}. Periodic boundary conditions are used because we are interested in bulk properties of the model. In simulations, we use the Metropolis Hastings algorithm with a local update scheme and parallel tempering between different temperatures to numerically evaluate various observables \cite{Katzgraber09, Press07, Newman99}. The gauge-field is  discretized through renormalized non-compact link-variables defined as
\begin{equation}
\label{eq:LattReg:LinkVarDef}
    A_{\v{r},\mu} \equiv -\frac{1}{g}\int_{\v{r}}^{\v{r} + \hat{\mu}}\!\!\!A_\mu(\v{r}')\;\mathrm{d}\v{r}' \in (-\infty, \infty),
\end{equation}
for $\mu\in\{x,y,z\}$. These are non-compact in the sense that they don't have a $2\pi$ periodicity \cite{Shimizu12} and this means that the discretization of the pure gauge term in Eq.~\eqref{eq:GL_kin+B} will have the form
\begin{equation}
\label{eq:LattReg:GaugeTermReg}
    \int\!\mathrm{d}^3r\;|\nabla\times\v{A}|^2\mapsto f_\text{A}^\text{r}=\frac{1}{g^2}\sum_{\v{r},\mu}(\v{\Delta}\times\v{A})^2,
\end{equation}
where \((\v{\Delta}\times\v{A})_\mu = \epsilon_{\mu\nu\lambda}\Delta_\nu A_{\v{r},\lambda}\) using the Levi-Civita symbol and summation over repeated indices. $\Delta_\mu$ is a discrete forward difference operator such that ${\Delta_\mu A_{\v{r},\nu} = A_{\v{r}+\hat{\mu},\nu} - A_{\v{r},\nu}}$. We note that writing out the sums over $\mu$, $\nu$ and $\lambda$, Eq.~\eqref{eq:LattReg:GaugeTermReg} can be written in term of plaquette sums. The link variables are renormalized in the sense that we multiply the field by a factor $-1/g$ to simplify the covariant derivatives.

The covariant derivatives are discretized using forward difference where the order-parameter component value at $\v{r}+\hat{\mu}$ is parallel-transported back to $\v{r}$ by the gauge-field link variables by
\begin{equation}
\label{eq:LattReg:CovDerReg}
    D_\mu\eta_a(\v{r})\mapsto\eta^a_{\v{r}+\hat{\mu}}e^{-iA_{\v{r},\mu}} - \eta^a_\v{r}.
\end{equation}
This ensures that the resulting lattice-discretized \ac{GL}-theory remains invariant under the gauge-transformation
\begin{equation}
\label{eq:LattReg:DiscreteGaugeTrans}
    \begin{split}
        \eta^a_\v{r}&\mapsto e^{i\lambda_\v{r}}\eta^a_\v{r}\\
        A_{\v{r},\mu}&\mapsto A_{\v{r},\mu} + \Delta_\mu\lambda_\v{r},
    \end{split}
\end{equation}
where $\lambda_\v{r}$ is an arbitrary real field.

The resulting lattice theory is expected to yield the same quantitative behaviour as the continuum theory, at least in strongly type-II regime \cite{Nguyen99}.
The remaining expressions for the discretized effective free energy density $f^\text{r}$ are presented in Appendix~\ref{app:FreeEnReg}. {Once a lattice formulation of the \ac{GL}-model is obtained, Monte Carlo simulations are carried out in the following manner. We start from some given configuration of the phase and gauge-field variables. In principle this configuration can be completely arbitrary, but we typically start from a fully correlated low temperature configuration or a fully uncorrelated high temperature configuration. From this configuration, we propose a local update by changing the field values at one lattice site. The update is either accepted or rejected, with a probability given by the Boltzmann weight of the change in energy between the old and proposed configuration. After identifying the two symmetry-sectors of the model, these updates are done in a two-step manner. We first attempt to change the phase-difference \(\theta_{\vect{r}}^x - \theta_{\vect{r}}^y\) while keeping the phase-sum constant. The second step is attempting to change the phase-sum \(\theta_{\vect{r}}^x + \theta_{\vect{r}}^y\) and the gauge field \(A_{\vect{r},\mu}\) while keeping the phase difference constant. All proposed updates were drawn from a uniform distribution about the value in the current configuration. For updates to the phase variables we set \(\theta_{max} = 2\pi/3\), such that the new value was drawn from the interval \([\theta_{\vect{r}} - \theta_{max}, \theta_{\vect{r}} + \theta_{max}]\) (mod \(2\pi\)). Similarly, updates to the gauge field were drawn from \([A_{\vect{r},\mu} - A_{max}, A_{\vect{r}, \mu} + A_{max}]\), where \(A_{max} = 0.6\). The values of \(\theta_{max}\) and \(A_{max}\) were set after initial testing to keep the acceptance rate around 30\%. This procedure respects the important requirements of ergodicity and detailed balance, and also has the benefit of allowing a high acceptance rate in one symmetry sector even if the other one is completely frozen. A Monte Carlo sweep consists of performing one such update on each lattice site in the system. Before doing measurements, we carry out a number of Monte Carlo sweeps to let the system thermalize at a configuration with high probability. This is done in a step-wise manner where the system is thermalized at incrementally higher or lower temperatures, depending on the starting configuration, towards the target temperature. This stepwise procedure decreases the probability of getting stuck in local minima of the energy landscape. Furthermore, our algorithm employs parallel tempering where a number of systems are running in parallel at an interval of closely spaced temperatures. After a number of Monte-Carlo sweeps, two configurations can swap temperature with a probability given by the Boltzmann weight. This further remedies issues associated with local minimas and ensures faster thermalisation and sampling.}

To measure ordering in each of the symmetry sectors at the phase transition, we introduce two order parameters. As discussed in section \ref{subsec:Charged_chiral_sym_sec}, the \(Z_2\) transition is characterized by an imbalance between the chiral components introduced in Eq.~\eqref{eq:chiral_amplitudes}. Hence, we can measure spontaneous symmetry-breaking of \ac{TRS} using the chiral amplitude difference
\begin{equation}\label{eq:chira_order_param}
    \delta \eta_{\pm} = \left\langle \left| \frac{1}{L^3}\sum_{\vect{r}} |\eta_+(\vect{r})|^2 - |\eta_{-}(\vect{r})|^2\right|\right\rangle.
\end{equation}
This is zero in the high-temperature phase and tends to \(2\rho_0^2\) in the low-temperature phase. The superconducting phase is characterized by a non-zero gauge field mass \(m = \lambda_L^{-1}\). This can be computed via  the dual stiffness \cite{PhysRevB.71.214509}
\begin{equation}\label{eq:dual_stiffness}
    \rho^{\mu\mu}_{\vect{q}} = \frac{1}{(2\pi)^2 L^3}\left\langle\left| \sum_{\vect{r}}(\Delta \times \vect{A})_\mu e^{i\vect{q}\vect{r}} \right| \right\rangle \sim \frac{q^2}{q^2 + \lambda_L^{-2}}.
\end{equation}
The low \(\vect{q}\)-limit of this expression tends to zero in the superconducting phase, where \(\lambda_L\) is finite, and some constant in the normal state, where \(\lambda_L\) is infinite. Hence, we measure the dual stiffness at the lowest non-zero momentum allowed by our discretization as an order parameter in the \(U(1)\)-symmetry sector. Finally, both phase-transitions are accompanied by singularities in the specific heat
\begin{equation}\label{eq:specific_heat}
    C_v = \beta^2\left\langle (E-\langle E \rangle)^2\right\rangle,
\end{equation}
where \(\beta\) is the inverse temperature. 

In numerical simulations, we thermalise systems of sizes up to \(32^3\) from both an ordered state given by Eqs.~\eqref{eq:MFGS_amps} and \eqref{eq:MFGS_phase}, or fully disordered states in some cases, over \(3\times10^5\) Monte Carlo sweeps. We then make measurements of the energy, dual stiffness and chiral order parameter over \(1\times10^6\) Monte Carlo sweeps. The measurements are done every 40'th sweep to account for the auto-correlation time. Ferrenberg-Swendsen multi histogram reweighting has been used to post-process the raw data \cite{PhysRevLett.61.2635, ferrenberg1989optimized}. {Errors in the results were estimated using the jackknife method \cite{Efron_1_1979}.} 
\section{Results} \label{sec:Results}
In this section we present results from large scale Monte Carlo simulations using the parameter regime discussed in Sec.~\ref{subsec:London_limit}. For all simulations we have fixed \(\alpha = 1.0\), \(u_0 = 1.0\), \(g=1.0\) {and we consider a cubic geometry for lattices of size \(L\times L\times L\) with periodic boundary conditions.}

\subsection{Model without mixed gradient terms}
Results without \ac{MGT}, \(\gamma_m = 0\), are shown in Fig.~\ref{fig:order_params_mgt0_nu0,1}. We find that ordering in both symmetry sectors occurs simultaneously. In Fig.~\ref{fig:chiral_order_mgt0_nu0,1} the chiral order parameter has a kink as it drops to zero at the critical temperature. The dual stiffness in Fig.~\ref{fig:dual_stiffness_mgt0_nu0,1} displays similar behaviour; in the Meissner phase, where \(\lambda_L\) is finite, it tends to zero and in the normal state it grows, as the thermal gauge fluctuations become larger. The normal phase and Meissner phase are separated by a jump in both order parameters accompanied by a singularity in the specific heat in Fig.~\ref{fig:specific_heat_mgt0_nu0,1}. In summary, we find that with decreasing temperature the system goes from a normal state to a chiral superconducting state with spontaneously broken \(U(1) \times Z_2\) symmetry.

\begin{figure}[htb]
    \subfloat[\label{fig:chiral_order_mgt0_nu0,1}]{
        \includegraphics[width=\linewidth]{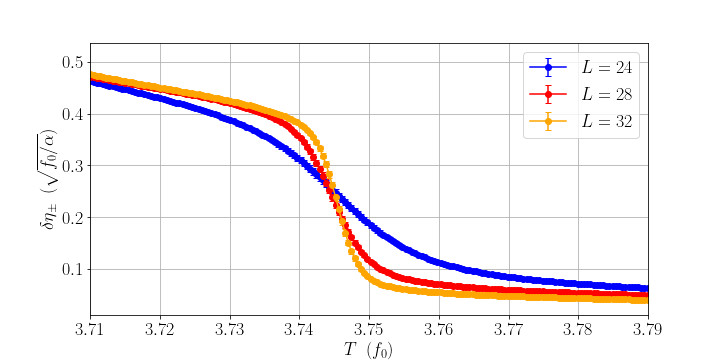}
    }
    
    \subfloat[\label{fig:dual_stiffness_mgt0_nu0,1}]{
        \includegraphics[width=\linewidth]{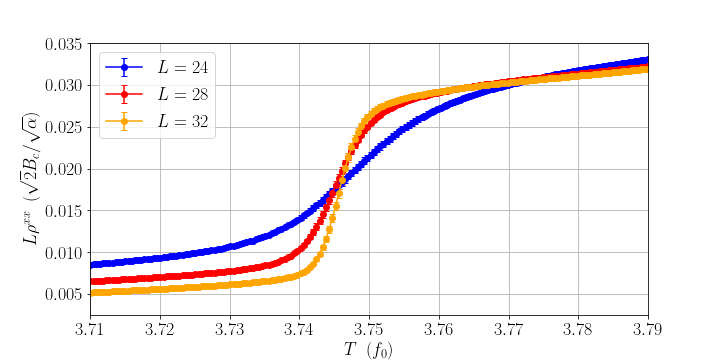}
    }
    
    \subfloat[\label{fig:specific_heat_mgt0_nu0,1}]{
        \includegraphics[width=\linewidth]{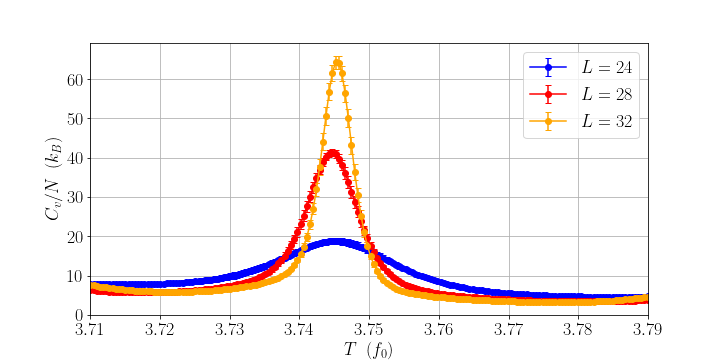}
    }
    \caption{Results from Monte Carlo simulations of model in Eq.~\eqref{eq:GL_full} with \(\gamma=0.1\) and \(\gamma_m = 0.0\) for \(L=24,\, 28, \,32\). (a) Chiral amplitude difference given by Eq.~\eqref{eq:chira_order_param}. (b) Dual stiffness given by Eq.~\eqref{eq:dual_stiffness}. (c) Specific heat given by Eq.~\eqref{eq:specific_heat}. We find a single phase transition at \(T_c \simeq 3.745\) characterized by ordering in both symmetry sectors and a singularity in the specific heat.}
    \label{fig:order_params_mgt0_nu0,1}
\end{figure}

\begin{figure}[htb]
\includegraphics[width=\linewidth]{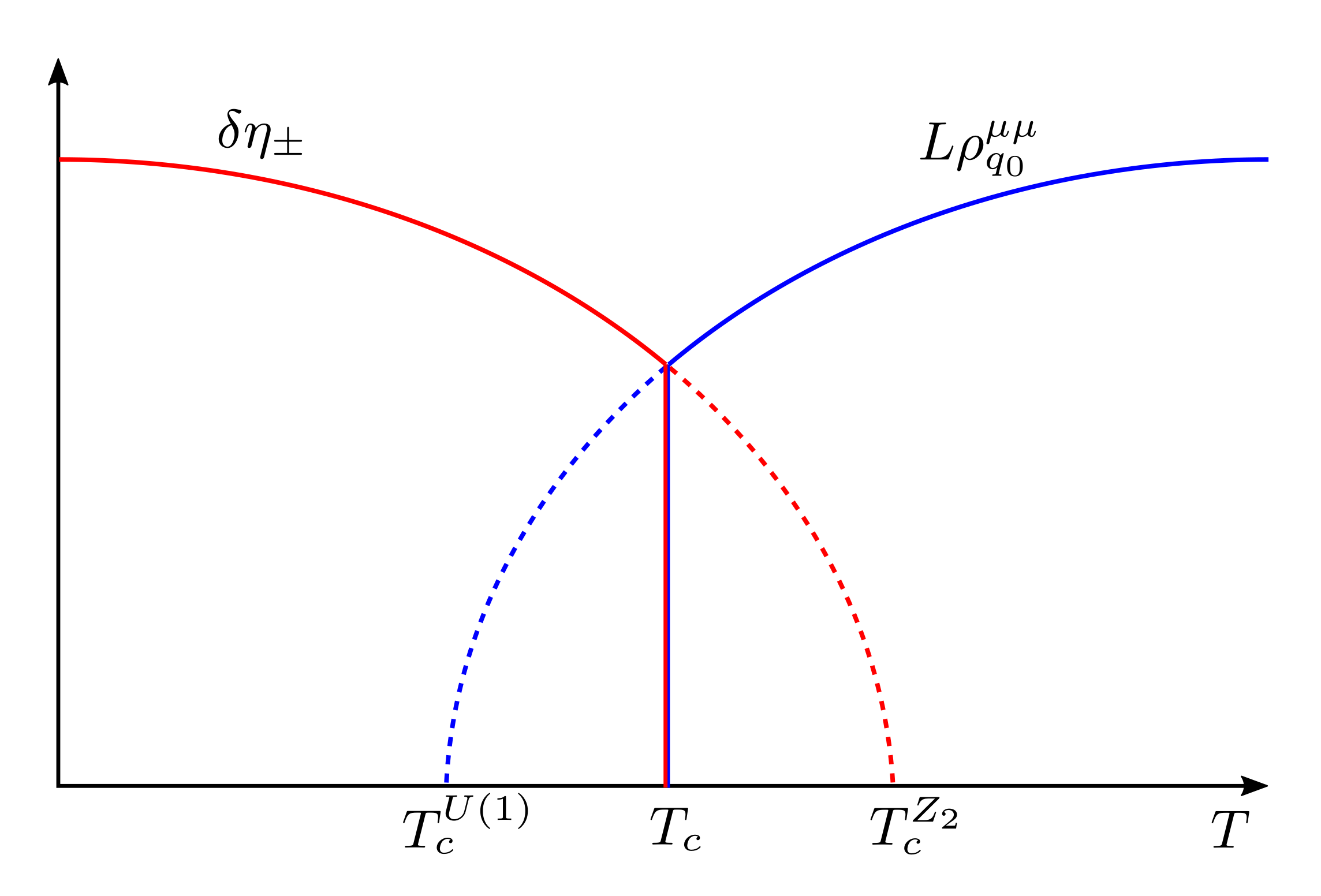}
\caption{Schematic drawing of the preemptive phase transition scenario. The chiral and charged order parameters would exhibit two separate continuous phase transitions, were it not for the mutual interplay between the two sectors. At some intermediate temperature, interplay between topological defects in the two symmetry sectors lead to ordering in both, resulting in a single first order phase transition at \(T_c\).}
\label{fig:Preemptive_phase_trans}
\end{figure}

The fact that they coincide is explained by a preemptive phase transition scenario, discussed previously for multi-component superfluids and superconductors \cite{PhysRevB.82.134511, dahl2008preemptive}, see the earlier discussion in terms of $j$-currents in \cite{kuklov2006deconfined}. The process of proliferating topological defects in the two symmetry sectors is cooperative. Namely, as the charged vortices in the \(U(1)\)-sector proliferate, the stiffness of the Ising domain walls drops to zero triggering a proliferation in the \(Z_2\) sector. The smoking gun signature of a preemptive phase transition is that it is first order, with a latent heat related to the sudden drop in the chiral/charged order parameters at the phase transition. An intuitive way of understanding this is to consider the case where the two symmetry sectors are completely decoupled. The chiral sector is then, with increasing temperature, headed towards a continuous second order phase transition in the Ising universality class. At some lower temperature, charged vortices in the \(U(1)\)-sector will proliferate which also triggers the \(Z_2\) phase transition due to the interplay between domain walls and vortices. 
This scenario is sketched in Fig.~\ref{fig:Preemptive_phase_trans}, where the order parameters in both symmetry sectors are cut off at the preemptive transition temperature resulting in a single first order transition. To investigate this numerically, we plot the energy probability distribution in Fig.~\ref{fig:Histograms_mgt0_nu0,1}. We find a pronounced double peak, indicative of a first order phase transition where two phases co-exist at the critical temperature. Furthermore, we have performed a finite-size scaling analysis of the difference in free energy between the double peak value and the valley minimum \(\Delta F = \ln(P_{max}/P_{min})/\beta\), where \(P_{max}\) and \(P_{min}\) are the energy probabilities at the double peak and the valley minimum, respectively.  For a first order phase transition, this quantity should scale asymptotically as \(L^{d-1}\) \cite{PhysRevLett.65.137} for large system sizes. Such scaling is confirmed in Fig.~\ref{fig:FSS_mgt0_nu0,1}. 

\begin{figure}[htb]
    \subfloat[\label{fig:Histograms_mgt0_nu0,1}]{
        \includegraphics[width=\linewidth]{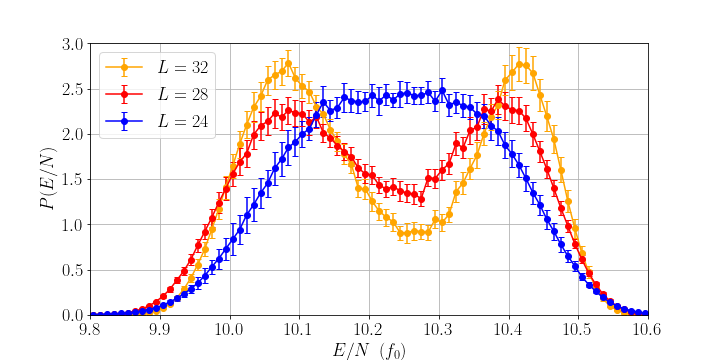}
    }
    
    \subfloat[\label{fig:FSS_mgt0_nu0,1}]{
        \includegraphics[width=\linewidth]{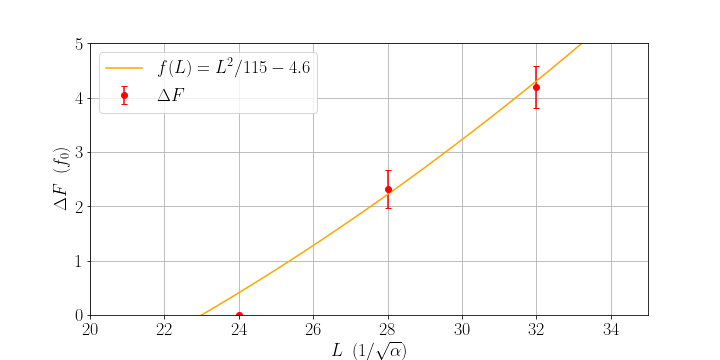}
    }
    \caption{(a) Energy per lattice site probability distribution at the critical temperature for system parameters \(\gamma = 0.1\) and \(\gamma_m = 0.0\) and system sizes \(L=24, 28, 32\). For larger system sizes we see an increasingly pronounced double peak, indicating a first order phase transition. (b) Finite size scaling of the difference in free energy between the double peak value \(P_{max}\) and the valley minimum \(P_{min}\), \(\Delta F = \ln(P_{max}/P_{min})/\beta\), measured at the critical point. Ferrenberg Swendsen multi histogram reweighting has been used to obtain histograms with peaks of similar height.}
    \label{fig:First_order_mgt0_nu0,1}
\end{figure}

\subsection{Full model}
We now consider the full model in Eq.~\eqref{eq:GL_full} and examine how the \ac{MGT} modify results from the previous section.
 Fig.~\ref{fig:order_params_mgt0,1_nu0,1} shows results for the phase transition at \(\gamma_m = 0.1\). The difference between the results with and  without \ac{MGT}, can be seen by comparing with Fig.~\ref{fig:order_params_mgt0_nu0,1}. The critical temperature decreases slightly, and we can also see that finite size effects become more prominent as the peak in specific heat changes more with system size. To investigate whether this is still a preemptive phase transition, the energy probability distribution along with finite size scaling of \(\Delta F\) are plotted in Fig.~\ref{fig:First_order_mgt0,1_nu0,1}. We find a clear double peak and quadratic scaling, which both indicate a first order preemptive phase transition. By comparing with Fig.~\ref{fig:First_order_mgt0_nu0,1},  we observe that the first order behavior is even stronger in case of non-zero \ac{MGT}, as the double peak structure is now resolved for the smallest system with \(L=24\). 
 {The discontinuous character of the phase transition is consistent with intuition from mean-field \ac{GL} solutions where vortices and domain walls tend to form a strong  bound states in chiral p-wave
superconductors \cite{ichioka2005magnetization,PhysRevB.94.104509,garaud2012skyrmionic}.} 

\begin{figure}[htb]
    \subfloat[\label{fig:chiral_order_mgt0,1_an0_nu0,1}]{
        \includegraphics[width=\linewidth]{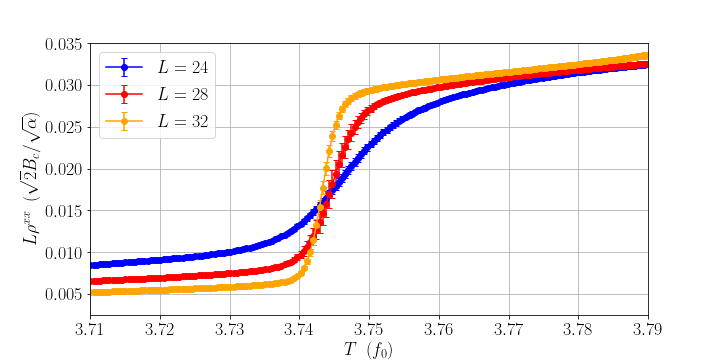}
    }
    
    \subfloat[\label{fig:dual_stiffness_mgt0,1_an0_nu0,1}]{
        \includegraphics[width=\linewidth]{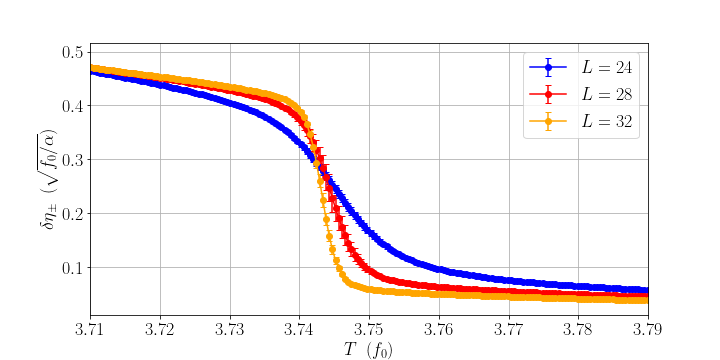}
    }
    
    \subfloat[\label{fig:specific_heat_mgt0,1_an0_nu0,1}]{
        \includegraphics[width=\linewidth]{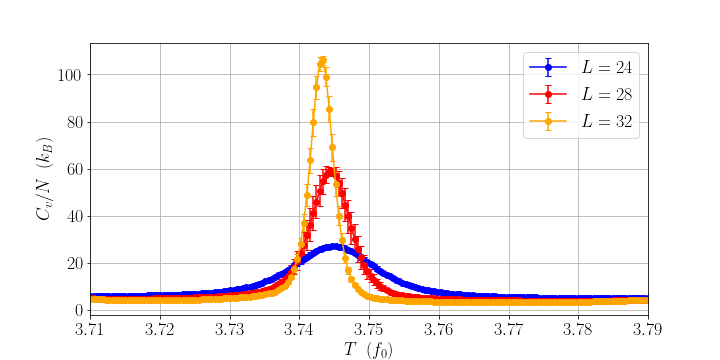}
    }
    \caption{Results from Monte Carlo simulations of model in Eq.~\eqref{eq:GL_full} with \(g=1.0\), \(\gamma=0.1\) and \(\gamma_m = 0.1\) for \(L=24,\, 28, \,32\). (a) Chiral amplitude difference given by Eq.~\eqref{eq:chira_order_param}. (b) Dual stiffness given by Eq.~\eqref{eq:dual_stiffness}. (c) Specific heat given by Eq.~\eqref{eq:specific_heat}. We find a single phase transition at \(T_c \simeq 3.743\) characterized by ordering in both symmetry sectors and a singularity in the specific heat.}
    \label{fig:order_params_mgt0,1_nu0,1}
\end{figure}

\begin{figure}[htb]
    \subfloat[\label{fig:Histograms_MGT0,1_nu0,1}]{
        \includegraphics[width=\linewidth]{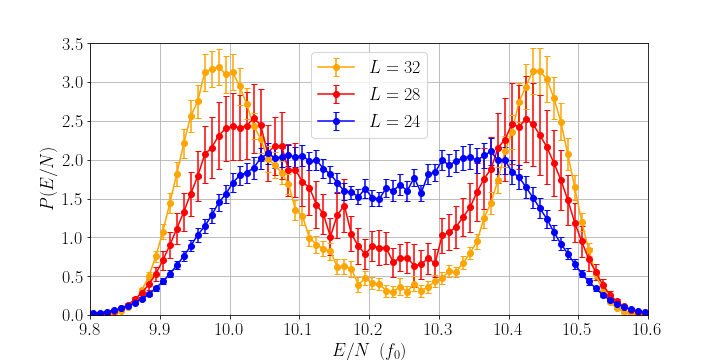}
    }
    
    \subfloat[\label{fig:FSS_MGT0,1_nu0,1}]{
        \includegraphics[width=\linewidth]{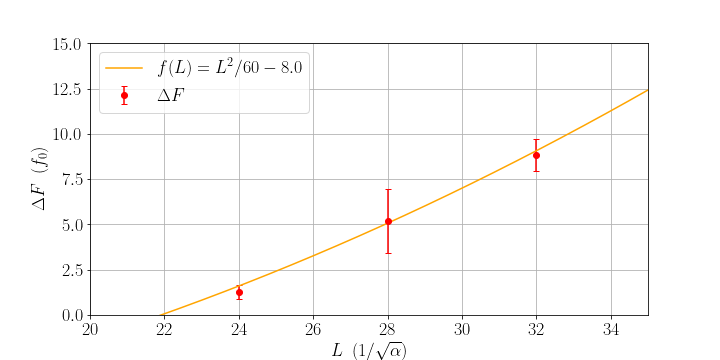}
    }
    \caption{(a) Energy per lattice site probability distribution at the critical temperature for system parameters \(\gamma = 0.1\) and \(\gamma_m = 0.1\) and system sizes \(L=24, 28, 32\). For larger system sizes we see an increasingly pronounced double peak, indicating a first order phase transition. (b) Finite size scaling of \(\Delta F\). Ferrenberg Swendsen multi histogram re-weighting has been used to obtain histograms with peaks of similar height.}
    \label{fig:First_order_mgt0,1_nu0,1}
\end{figure}

\begin{table}[htb]
    \centering
    \begin{tabular}{p{4em}c c c}
        \hline
         \(\gamma_m\) & \(T_c\) & \(c\) \\  
         \hline
         0.0 & \(3.7452 \pm 0.0004\) & \(0.088\pm0.004\) \\
         0.1 & \(3.7434 \pm 0.0004\) & \(0.120\pm0.004\) \\
         0.2 & \(3.7359 \pm 0.0004\) &  \(0.121\pm0.004\) \\
         \hline
    \end{tabular}
    \caption{Critical temperature \(T_c\) and coefficient for the change in entropy \(\Delta S = c k_B\) for different strengths of the \ac{MGT} with \(\gamma=0\). As \(\gamma_m\) increases the critical temperature decreases and the change in entropy increases, making the phase transition stronger first order. Data is taken from simulations with \(L=32\), and \(T_c\) is determined using multi histogram re-weighting to find the temperature where the two peaks in the energy probability distribution have the same height. {The uncertainty in \(T_c\) and \(c\) are determined by our numerical resolution in energy and temperature.}} 
    \label{tab:entropy_difference}
\end{table}

To characterize the strength of the transition, we calculate the difference in entropy between the two coexisting states at the phase transition. The entropy is calculated from the free energy \(F = E – TS\). Because the two states have the same free energy the entropy difference is given by \(\Delta S = \Delta E/T_c \equiv c k_b\). In Table \ref{tab:entropy_difference} we show the coefficient \(c\) for the change in entropy per lattice site for increasing values of \(\gamma_m\). We see a significant increase from the case without to the case with \ac{MGT}, meaning the phase transition becomes more strongly first order. As \(\gamma_m\) is increased further, this trend continues. 
This can be explained by the fact that the  \ac{MGT} introduce stronger interaction between vortices and domain walls that results in 
 a larger latent heat and stronger first order behaviour. 

\section{Summary and discussion}
In this paper we have investigated fluctuation effects on the phase transition in a \ac{GL} model for chiral superconductors. Within the parameter regime we have used, consistent with taking $(\eta_x,\eta_y)$ to be constants, a single phase transition from the normal state to a chiral superconducting state with spontaneously broken \(U(1)\times Z_2\)-symmetry is found. We show that this is a preemptive first-order phase transition, where interplay between the topological defects in both symmetry sectors of the model cause them both to disorder at the same temperature. We have also investigated the effect of \ac{MGT}-term, which enhance the first order character of the phase transition.

An issue that we have not dealt with in this paper, is whether we can tune parameters of the model such that the discontinuous  phase transition we find may be separated into $U(1)$ and $Z_2$.
Such a separation was demonstrated in $s+is$ models \cite{Bojesen2013,Bojesen2014,grinenko2021bosonic} and was discussed also in a chiral p-wave model \cite{fischer2016fluctuation}. In principle, one can increase the critical temperature of the charged sector alone by lowering the gauge charge \(g\) and similarly decrease the critical temperature of the chiral sector by lowering \(\gamma\) in an attempt to swap the order of the two phase transitions (i.e. make the critical temperature of the $Z_2$ transition smaller than that of the $U(1)$ transition). However, Eq.~\ref{eq:Sep_of_variables} shows that in the limit \(g=\gamma=0\) (ignoring \ac{MGT}) both symmetry sectors are reduced to global \(U(1)\). Since these have the same stiffness, \(\rho_0/2\), they will also have the same critical temperature, meaning the two phase-transitions can never swap place in the considered model. Additionally, the \ac{MGT} in Eq (\ref{eq:GL_full}) make vortices more strongly bound to domain walls than in the absence of \ac{MGT}. However, adding different \ac{MGT} fourth order in fields and second order in gradients like in \cite{grinenko2021bosonic} should produce splitting as it tunes  domain wall energy relative to the vortex energy.

{The results presented in this paper are relevant for
superconductors with spontaneously broken time-reversal symmetry and a \(D_{4h}\) symmetry group, that can be described by the energy density in Eq.~\eqref{eq:GL_full}. In general, spin-triplet superconductors with Fermi-surface anisotropy, Fermi-level particle-hole anisotropy, or spin-orbit coupling, will yield a \ac{GL} theory of the type we have used in this work. The parameter regime considered is limited by the fixed amplitude approximation discussed in Sec.~\ref{subsec:London_limit}, which also puts clear restrictions on real materials that could display the behaviour found in this paper. The main merit of the paper is therefore insight into the nature of the phase transition in chiral \(p\)-wave superconductors in a parameter regime where phase and gauge field fluctuations are dominant, {that can be expected in strongly type-II regime \cite{PhysRevLett.47.1556}}.}

\section{Acknowledgements}
We acknowledge financial support from the Research Council of Norway Grant No. 262633 “Center of Excellence on Quantum
Spintronics,” and Grant No. 250985, “Fundamentals of Low Dissipative Topological Matter.”
EB was supported by the Swedish Research Council Grants  2016-06122, 2018-03659, the G\"{o}ran Gustafsson Foundation for Research in Natural Sciences and Medicine.


\bibliography{main.bib}

\appendix
\section{Coupling matrix}\label{app:Coupling_matrix}
The energy in Eq.~\eqref{eq:GL_full} is expanded to second order in the fluctuation fields introduced in Eqs.~\eqref{eq:epsilon_i}-\eqref{eq:p_i}. For the potential in Eq.~\eqref{eq:GL_pot}, we find
\begin{equation}\label{eq:fp_two_comp_isotrop}
\begin{split}
    f_\mathrm{V} = & -\alpha\left(u_x^2 + u_y^2\right) + \frac{u_0}{2}\left(u_x^4 + u_y^4\right) - \gamma u_x^2u_y^2 \\
    & + \left(-\alpha + 3u_0 u_x^2 - \gamma u_y^2\right)\epsilon_x^2 + \left(-\alpha + 3u_0 u_y^2 - \gamma u_x^2\right)\epsilon_y^2 \\
    & - 4\gamma u_x u_y \epsilon_x \epsilon_y + 8\gamma u_x^2 u_y^2 \theta_\Delta^2,
\end{split}
\end{equation}
where \(f_0\) is the ground state energy. The Maxwell term keeps the exact same form to second order in \(\vect{p}\), since it only differs by a gradient from \(\vect{A}\)
\begin{equation}
    f_\mathrm{A} = |\nabla \times \vect{p}|^2.
\end{equation}
Finally for the various gradient terms, we expand the gradients to first order since all relevant combinations are squared
\begin{alignat}{1}
    D_i\eta_x & = \left[\partial_i \epsilon_x - i(gp_i - \partial_i\theta_\Delta)u_x\right]e^{i(\theta_\Sigma + \theta_0/2)}, \\
    D_i\eta_y & = \left[\partial_i \epsilon_y - i(gp_i + \partial_i\theta_\Delta)u_y\right]e^{i(\theta_\Sigma - \theta_0/2)}.
\end{alignat}
These expressions can now be combined to form all the terms in the energy functional. After Fourier transforming and rotating the amplitude basis, we can write the energy on the form in Eq.~\eqref{eq:GL_energy_matrix}
\begin{equation}
    f = f_0 + \vect{v}\vect{G}\vect{v}^\dagger.
\end{equation}
where \(\vect{v}\) is given in Eq.~\eqref{eq:fluctuation_vector}. We can write the coupling matrix as a sum of three contributions, a diagonal part with massive terms, a diagonal part with \(k\)-dependant terms and an off-diagonal part from the \ac{MGT}.
\begin{equation}
    \vect{G} = \vect{G}_{\mathrm{D}}  + \vect{G}_{\mathrm{MGT}}
\end{equation}
\begin{widetext}
\begin{equation}\label{eq:G_d0}
    \vect{G}_{\mathrm{D}} = \begin{pmatrix}
    \frac{2\alpha(u_0+\gamma)}{u_0-\gamma} + k^2 & 0 &0 &0 &0 \\
    0 & 2\alpha + k^2 & 0 & 0 & 0 \\
    0 & 0 & \frac{8\gamma\alpha^2}{(u_0-\gamma)^2} + \frac{2\alpha}{u_0-\gamma}k^2& 0 & 0 \\
    0 & 0 & 0 & \frac{2\alpha}{u_0-\gamma}g^2 +k^2 & 0 \\
    0 & 0 & 0 & 0 & \frac{2\alpha}{u_0-\gamma}g^2 +k^2
    \end{pmatrix}
\end{equation}
\begin{equation}\label{eq:G_mgt}
    \vect{G}_{\mathrm{MGT}} = \pm\gamma_m\sqrt{\frac{2\alpha}{u_0-\gamma}} \begin{pmatrix}
    0 & 0 & - 2 k_x k_y & 0 & 0 \\
    0 & 0 & 0 & - i g  k_y & - i  g  k_x \\
    - 2 k_x k_y & 0 & 0 & 0 & 0 \\
    0 &  i  g  k_y & 0 & 0 & 0 \\
    0 &  i  g  k_x & 0 & 0 & 0 
    \end{pmatrix}
\end{equation}
\end{widetext}
In Eq.~\eqref{eq:G_d0}, we note that the Meissner effect gives rise to massive gauge-field fluctuations, which yield a massless Goldstone mode associated with the phase-sum when  \(g =0\). The phase-difference mode is also seen to evolve to a massless Goldstone mode when the Ising-anisotropy parameter $\gamma=0$.
Furthermore, Eq.~\eqref{eq:G_mgt} shows that the \ac{MGT} have an effect for \(g=0\), coupling fluctuations in the \(\epsilon_+\) amplitude mode to fluctuations in the phase difference \(\theta_\Delta\). Finite \( g \)
will moreover couple the \(\epsilon_-\)  amplitude mode   to gauge-invariant currents. Contrary to the one-component case, the eigenmodes are in general complicated linear combinations of amplitude modes, phase-difference modes, and gauge-invariant currents \cite{speight2019chiral}. Only in a limited parameter regime do the eigenmodes simplify significantly. 
 
\section{Lattice regularized free energy}
\label{app:FreeEnReg}
In this section we apply the regularization procedure introduced in Section~\ref{sec:MC_sim} to the dimensionless effective free energy density in Eq.~\eqref{eq:GL_full}. The resulting expression was used in the Metropolis-Hastings algorithm to find the energy-difference between different field-configurations as well as when calculating the energy as an observable which again was used in calculating of the specific heat.

Inserting the discretization of the covariant derivative in Eq.~\eqref{eq:LattReg:CovDerReg} yields
\begin{equation}
    \begin{split}
        |D_\mu\eta_a|^2 \mapsto &|\eta^a_{\v{r}+\hat{\mu}}|^2 + |\eta^a_\v{r}|^2 - 2\Re\big(\eta^a_{\v{r}+\hat{\mu}}\eta^{a\;\ast}_\v{r}e^{-iA_{\v{r},\mu}}\big)\\
        \sim&2\Big[(\rho^a_\v{r})^2 - \rho^a_{\v{r}+\hat{\mu}}\rho^a_\v{r}\cos(\theta^a_{\v{r}+\hat{\mu}} - \theta^a_\v{r} - A_{\v{r},\mu})\Big].
    \end{split}
\end{equation}
In the second line we have introduced the notation ${\eta^a_\v{r} = \rho^a_\v{r}e^{i\theta^a_\v{r}}}$ for the amplitude and phase of the components of the order parameter. We have also used periodic boundary conditions to map the term $|\eta^a_{\v{r}+\hat{\mu}}|^2$ back to $|\eta^a_\v{r}|^2$ by a simple shift of the index in the sum $\sum_\v{r}f_\v{r}$.

Using the formula above we immediately get the lattice-regularized conventional kinetic energy density
\begin{equation}
    \begin{split}
        f_\text{K}^\text{r} &= \reg\Big\{\sum_a|\v{D}\eta_a|^2\Big\} = \sum_{\mu\,a}\reg\big\{|D_\mu\eta_a|^2\big\}\\
        &= 2\sum_{\mu\,a}\Big[(\rho^a_\v{r})^2 - \rho^a_{\v{r}+\hat{\mu}}\rho^a_\v{r}\cos\big(\theta^a_{\v{r}+\hat{\mu}} - \theta^a_\v{r} - A_{\v{r},\mu}\big)\Big],
    \end{split}
\end{equation}
where $\mu$ runs over $x$, $y$ and $z$, while $a\in\{x,y\}$.
Using the notation 
\begin{equation}
    \bar{a} = \left\{
    \begin{array}{lr}
         y &: a=x  \\
         x &: a=y
    \end{array}\right..
\end{equation}
the \ac{MGT} in Eq.~\eqref{eq:GL_mgt} can be written on the more compact form
\begin{equation}
    f_\text{MGT} = 2\gamma_m\sum_a\Re\Big[D_x\eta_a(D_y\eta_{\bar{a}})^\ast\Big].
\end{equation}
Inserting the discretization of covariant derivatives we find in Eq.~\eqref{eq:LattReg:CovDerReg} gives
\begin{alignat}{1}
    D_x\eta_a(D_y\eta_{\bar{a}})^{\ast} = &  \left(\rho_{\vect{r}+\hat{x}}^a e^{i(\theta_{\vect{r}+\hat{x}}^a - A_{\vect{r},x})} -\rho_{\vect{r}}^a e^{i\theta_{\vect{r}}^a}\right)\nonumber \\
    & \times \left(\rho_{\vect{r}+\hat{y}}^{\bar{a}} e^{-i(\theta_{\vect{r}+\hat{y}}^{\bar{a}} - A_{\vect{r},y})} -\rho_{\vect{r}}^{\bar{a}} e^{-i\theta_{\vect{r}}^{\bar{a}}}\right)\nonumber \\
    & = \rho_{\vect{r}+\hat{x}}^a\rho_{\vect{r}+\hat{y}}^{\bar{a}}e^{i(\theta_{\vect{r}+\hat{x}}^a - \theta_{\vect{r}+\hat{y}}^{\bar{a}}-(A_{\vect{r},x} -A_{\vect{r}, y}))} \nonumber \\
    & -\rho_{\vect{r}+\hat{x}}^a\rho_{\vect{r}}^{\bar{a}}e^{i(\theta_{\vect{r}+\hat{x}}^a - \theta_{\vect{r}}^{\bar{a}}-A_{\vect{r},x})} \nonumber \\
    & -\rho_{\vect{r}}^a\rho_{\vect{r}+\hat{y}}^{\bar{a}}e^{-i(\theta_{\vect{r}+\hat{y}}^{\bar{a}}-\theta_{\vect{r}}^a - A_{\vect{r},y})} \nonumber \\
    & +\rho_{\vect{r}}^a\rho_{\vect{r}}^{\bar{a}}e^{i(\theta_{\vect{r}}^a - \theta_{\vect{r}}^{\bar{a}})}
\end{alignat}
Taking the real part of this gives
\begin{alignat}{1}
        \Re&\Big[D_x\eta_a(D_y\eta_{\bar{a}})^\ast\Big] \nonumber \\
        & =\rho_{\vect{r}+\hat{x}}^a\rho_{\vect{r}+\hat{y}}^{\bar{a}}\cos\left(\theta_{\vect{r}+\hat{x}}^a - \theta_{\vect{r}+\hat{y}}^{\bar{a}}-(A_{\vect{r},x}-A_{\vect{r}, y})\right) \nonumber \\
        & -\rho_{\vect{r}+\hat{x}}^a\rho_{\vect{r}}^{\bar{a}} \cos\left(\theta_{\vect{r}+\hat{x}}^a - \theta_{\vect{r}}^{\bar{a}}-A_{\vect{r},x}\right) \nonumber \\
        & -\rho_{\vect{r}+\hat{y}}^{\bar{a}}\rho_{\vect{r}}^{a}\cos\left(\theta_{\vect{r}+\hat{y}}^{\bar{a}}-\theta_{\vect{r}}^{a} - A_{\vect{r},y} \right) \nonumber \\
        & +\rho_{\vect{r}}^a\rho_{\vect{r}}^{\bar{a}}\cos\left(\theta_{\vect{r}}^a - \theta_{\vect{r}}^{\bar{a}}\right),
\end{alignat}
This gives the final expression for the discretized \ac{MGT}
\begin{equation}
    \begin{split}
        f&_\text{MGT}^\text{r} = 2\gamma_m\sum_a\Big[\rho^a_\v{r}\rho^{\bar{a}}_\v{r}\cos\big(\theta^a_\v{r}-\theta^{\bar{a}}_\v{r}\big)\\
        & - \rho^a_{\v{r}+\hat{x}}\rho^{\bar{a}}_\v{r}\cos\big(\theta^a_{\v{r}+\hat{x}} - \theta^{\bar{a}}_\v{r}-A_{\v{r},x}\big) \\
        & - \rho^a_{\v{r}+\hat{y}}\rho^{\bar{a}}_\v{r}\cos\big(\theta^a_{\v{r}+\hat{y}} - \theta^{\bar{a}}_\v{r}-A_{\v{r},y}\big) \\
        &  + \rho^a_{\v{r}+\hat{x}}\rho^{\bar{a}}_{\v{r}+\hat{y}}\cos\big(\theta^a_{\v{r}+\hat{x}} - \theta^{\bar{a}}_{\v{r}+\hat{y}}- (A_{\v{r},x}-A_{\v{r},y})\big) \Big],
    \end{split}
\end{equation}
where we have switched the superscripts \(a\leftrightarrow\bar{a}\) on the third line. To ensure that this discretized term is rendered invariant under the four-fold rotations of the square numerical lattice, we may average as follows
\begin{align}
f&_\text{MGT}^\text{r} \to \tilde{f}_\text{MGT}^\text{r} \\
= &\frac{1}{4}
\left[f_\text{MGT}^\text{r} +C_{4}f_\text{MGT}^\text{r} +C_4^2f_\text{MGT}^\text{r}+C_4^3 f_\text{MGT}^\text{r}
\right]
\end{align}
where $C_4$ denotes a 90 degree counterclockwise rotation of the xy-coordinate system. We then find
\begin{equation}
    \begin{split}
        \tilde{f}&_\text{MGT}^\text{r}= \frac{\gamma_m}{2}
        \sum_a\Big[\rho^a_{\v{r}+\hat{x}}\rho^{\bar{a}}_{\v{r}+\hat{y}}\cos\big(\theta^a_{\v{r}+\hat{x}} - \theta^{\bar{a}}_{\v{r}+\hat{y}}- (A_{\v{r},x}-A_{\v{r},y})\big)\\
        & - \rho^a_{\v{r}-\hat{x}}\rho^{\bar{a}}_{\v{r}+\hat{y}}\cos\big(\theta^a_{\v{r}-\hat{x}} - \theta^{\bar{a}}_{\v{r}+\hat{y}}+ (A_{\v{r}-\hat{x},x}+A_{\v{r},y})\big) \\
        & + \rho^a_{\v{r}-\hat{x}}\rho^{\bar{a}}_{\v{r}-\hat{y}}\cos\big(\theta^a_{\v{r}-\hat{x}} - \theta^{\bar{a}}_{\v{r}-\hat{y}}- (A_{\v{r}-\hat{y},y}-A_{\v{r}-\hat{x},x})\big) \\
        &  - \rho^a_{\v{r}+\hat{x}}\rho^{\bar{a}}_{\v{r}-\hat{y}}\cos\big(\theta^a_{\v{r}+\hat{x}} - \theta^{\bar{a}}_{\v{r}-\hat{y}}- (A_{\v{r},x}+A_{\v{r}-\hat{y},y})\big) \Big],
    \end{split}
\end{equation}

The potential terms in Eq.~\eqref{eq:GL_full} are simply discretized by mapping to the amplitude phase-notation and become
\begin{equation}
        f_\text{V}^\text{r} = \sum_a\Big[-\alpha(\rho^a_\v{r})^2 + \frac{u_0}{2}(\rho^a_\v{r})^4\Big] + \gamma\,(\rho^x_\v{r}\rho^y_\v{r})^2\cos 2 (\theta^x_\v{r} - \theta^y_\v{r}).
\end{equation}
These expressions together with the regularization of the pure gauge-potential term in Eq.~\eqref{eq:LattReg:GaugeTermReg} then give the complete discretized free energy density 
\begin{equation}
    f^\text{r} =  f^\text{r}_\text{V} + f^\text{r}_\text{K} + \tilde{f}^\text{r}_\text{MGT} + f^\text{r}_\text{A}.
\end{equation}

\end{document}